# Topological insulator Bi$_2$Se$_3$ thin films grown on double-layer graphene by molecular beam epitaxy

Can-Li Song[1,2], Yi-Lin Wang[1], Ye-Ping Jiang[1,2], Yi Zhang[1], Cui-Zu Chang[1,2], Lili Wang[1], Ke He[1], Xi Chen[2], Jin-Feng Jia[2], Yayu Wang[2], Zhong Fang[1], Xi Dai[1], Xin-Cheng Xie[1], Xiao-Liang Qi[3], Shou-Cheng Zhang[3], Qi-Kun Xue[1,2*], and Xucun Ma[1*]

[1] *Institute of Physics, Chinese Academy of Sciences, Beijing 100190, China*

[2] *Department of Physics, Tsinghua University, Beijing 100084, China*

[3] *Department of Physics, Stanford University, Stanford, California 94305-4045, USA*

**Atomically flat thin films of topological insulator Bi$_2$Se$_3$ have been grown on double-layer graphene formed on 6H-SiC(0001) substrate by molecular beam epitaxy. By a combined study of reflection high energy electron diffraction and scanning tunneling microscopy, we identified the Se-rich condition and temperature criterion for layer-by-layer growth of epitaxial Bi$_2$Se$_3$ films. The as-grown films without doping exhibit a low defect density of** $1.0\pm0.2\times10^{11}$/cm$^2$**, and become a bulk insulator at a thickness of 10 quintuple layers, as revealed by** *in situ* **angle resolved photoemission spectroscopy measurement.**

Three dimensional (3D) topological insulators (TIs) such as $Bi_2Se_3$ and $Bi_2Te_3$ are characterized by a bulk energy gap of strong spin-orbit coupling origin and metallic surface states protected by time-reversal symmetry.[1-11] Without external doping, the Fermi level ($E_F$) of the materials resides in the bulk energy gap and thus only intersects the Dirac cone topological surface states,[1-6] which we call as "intrinsic" TI hereafter. A major experimental breakthrough in the emerging field of TIs is the material realization of high-quality single crystals by self-flux method.[2,3,5,12] However, as revealed by angle resolved photoemission spectroscopy (ARPES), the as-grown samples without doping are not a bulk insulator, rather, an n-type doped semiconductor,[2,3,5,12] presumably due to Se (Te) vacancies and antisite defects.[12-14] Further improvement in material quality via reducing/removing the Se (Te) vacancies is limited: it requires significant bulk diffusion of Se (Te) atoms in the quartz tube where the crystal forms, which is difficult because of the very volatile nature of Se or Te molecules compared to Bi. To reduce the unwanted electron carriers and move the $E_F$ into the gap, a great number of carriers of opposite sign were introduced by extremely heavy doping with Sn atoms in $Bi_2Te_3$[5] and Ca atoms in $Bi_2Se_3$.[3,12]

In this work, we report molecular beam epitaxy (MBE) growth of high-quality $Bi_2Se_3$ thin films on double-layer graphene formed on the commercial 6H-SiC(0001) substrates. The double-layer graphene has a similar layered structure to $Bi_2Se_3$ and is chemically inert due to the strong bonding of carbon atoms,[15] which turns out to greatly suppress interface reaction leading to atomically sharp interface between $Bi_2Se_3$ and graphene. We identify the layer-by-layer growth conditions of $Bi_2Se_3$ immediately on graphene by real-time reflection high energy electron diffraction (RHEED). *In situ* scanning tunneling microscope (STM) and ARPES measurements show that the as-grown films by our method are indeed a bulk insulator when the thickness is above 10 quintuple layers (QL). The resulting sharp interface can be employed to study the interaction of the two distinct massless Dirac fermions in TI and graphene, respectively.

Our MBE growth and characterization of $Bi_2Se_3$ thin films were conducted in an ultrahigh vacuum (UHV) MBE-ARPES-STM combined system (Omicron) with a base pressure of $1\times10^{-10}$ Torr. The double-layer graphene was prepared on a nitrogen-doped SiC(0001) substrate with a resistivity of ~0.1 Ω·cm using the well-established recipe.[16] High purity Bi (99.999%) and Se (99.999%) were evaporated from standard Knudsen-cells. ARPES spectra were collected at room temperature by a Gammadata Scienta SES-2002 analyzer with a HeI light source of 21.2eV. We also used a low temperature STM system to characterize the as-grown films (Unisoku).[17] An MBE chamber is attached to the STM system and $Bi_2Se_3$ films were grown by using the same conditions mentioned above. STM images were taken at 4.8 K at a constant current of 0.1nA with a polycrystalline W tip. The *dI/dV* mapping was acquired using a lock-in technique with a bias modulation of 10mV at 987.5Hz.[17]

The optimal conditions for layer-by-layer growth of bulk insulating $Bi_2Se_3$ films were established by a systematic investigation of the growth dynamics, surface morphology and thickness-dependent electronic structure with RHEED, STM and ARPES. Similar to those for GaAs[18] and $Bi_2Te_3$,[6] the optimal growth can be achieved under Se-rich atmosphere (Se/Bi beam flux ratio θ ≥10) and $T_{Bi}>T_{substrate}>T_{Se}$ ($T_{Bi}$, $T_{substrate}$ and $T_{Se}$ are the Bi-cell, Se-cell and substrate temperatures, respectively). The former not only assures nucleation of the $Bi_2Se_3$ phase than other phases, according to the phase diagram,[19] but also minimizes the formation of Se vacancy, which is difficult in self-flux method.[2,3,5,12] Under Se-rich condition, there always exist extra Se molecules on the growing front. However, these molecules cannot be incorporated into the film and will desorb since $T_{substrate}>T_{Se}$. The optimal growth is demonstrated in Fig. 1. The film shown in this figure was grown at θ=11, $T_{Bi}$=550°C, $T_{Se}$=136°C, $T_{substrate}$=220°C and has a nominal thickness of 26 QL. The characteristic RHEED intensity oscillation recorded during film deposition indicates an ideal layer-by-layer growth (Fig. 1a), where each circle of the oscillations corresponds to deposition of one QL of $Bi_2Se_3$. Therefore, the growth rate can be precisely measured, which is ~0.3 QL/min in this case. The atomically flat nature of the film is first evidenced by the sharp 1×1 RHEED pattern in Fig. 1b and further demonstrated by the STM image in Fig. 1c. As expected, the steps seen in the STM image have a height of 0.95 nm, exactly the QL unit cell size of $Bi_2Se_3$ along the c axis.[20] The Se-rich condition results in very few defects in the as-grown samples. A typical STM image of the film is shown in Fig. 1d. Only two defects can be found in an area of 50×50 $nm^2$. The dominant defects are Se vacancies (the dark depressions in Fig. 1d), which can be identified by their registry with respect to the topmost Se lattice. The areal defect density in this sample is ~1.0±0.2×$10^{11}$/$cm^2$, which is much lower than that on the cleaved samples.[12] Further investigation indicates fewer vacancy defects if higher Se/Bi flux ratio is used.

The electronic structure of the film is studied by ARPES and shown in Fig. 2a. The Dirac cone (the dashed red lines) can be clearly seen, and the Dirac point is located at 130 meV below $E_F$. While the position of the Dirac point can change (several tens meV) with film thickness, what's important here is that the intense electronic pocket from the bulk conduction band observed in the cleaved crystals[2,3,12] does not appear, suggesting that the film at this thickness is already a bulk insulator. Given the conditions we used, it is found that the minimum thickness for the film being an "intrinsic" TI is 11 QL. The details of this result has been discussed elsewhere.[4]

We also tried another substrate Si(111)-7×7. To avoid possible formation of $SiSe_2$ and other selenides at the interface,[21] the Bi-terminated √3×√3 structure was first grown.[20] In spite of it, the morphology of the $Bi_2Se_3$ films on the Bi-terminated √3×√3 surface, especially in the very thinner regime, is ill-defined. Moreover, the films suffer from stronger electron-doping, which can be seen by a comparison of the ARPES data on both substrates (Fig. 2b). Clearly, the $Bi_2Se_3$(111) films on graphene show much quicker development to the bulk insulating state than those on the √3×√3-Bi/Si(111) substrate, suggesting more defects and poorer film quality in the latter case. The critical thickness for "intrinsic" TIs on the Bi-terminated Si(111) surface is as high as 24 QL.

Now we discuss the role of the double-layer graphene. In a heteroepitaxial growth, the interface plays very critical roles in growth mode, strain relaxation and formation of defects such as misfit and screw dislocations. The situation becomes even more severe when a film is being deposited on a substrate of different material, for example, $Bi_2Se_3$ on Si in which strong surface selenidization by Se molecules may occur. By using inert graphene substrate, we found that not only is the interface reaction completely suppressed, but also could the strain relaxation be complete at the very first epitaxial layer of $Bi_2Se_3$, which doesn't necessarily induce any dislocations expected from the large lattice mismatch between graphene and $Bi_2Se_3$. Figures 3a and 3b show the morphology and hexagonal atomic lattice of the graphene, respectively, prepared on 6H-SiC(0001). After ~1.5 QL deposition, single crystalline $Bi_2Se_3$ film is formed (Fig. 3c). At this stage, both 1 QL and 2 QL islands are observed, implying an incomplete wetting of the graphene surface. From the atomically resolved STM image of the 1 QL $Bi_2Se_3$ film in Fig. 3d, we know that $Bi_2Se_3$(111)-1×1 surface is formed. The surface lattice constant measured with respect to the exposed graphene is 0.41±0.01 nm, consistent with that (0.414 nm) of the bulk $Bi_2Se_3$(111) surface. The observation reveals that the strain is already fully relaxed even at the first QL. According to Figs. 3b and 3d, the close-packed direction of the Se atoms is rotated by 28° with respect to that of the graphene. This leads to a well-defined Morié pattern with a period of ~7.10 nm, as observed from *dI/dV* mapping at -0.35 V (Fig. 3e). Assuming that the $Bi_2Se_3$(111) film is fully relaxed, a simple simulation (Fig. 3f) indicates that, to form the Morié pattern in Fig. 3e, the in-plane close-packed direction of the $Bi_2Se_3$(111) film should be rotated by 29°, which is very close to the experimental value. It further supports a fully relaxed $Bi_2Se_3$ film at 1 QL. Evidently, no interface reaction occurs and the $Bi_2Se_3$/graphene interface is atomically sharp.

In summary, we have shown that graphene can serve a superb substrate for growing high quality $Bi_2Se_3$ films. Under the growth criteria established in this study, intrinsic TI films can be easily obtained when the thickness is above 10 QL. We emphasize that the growth conditions presented here can be applied to growth of other V-VI topological thin films and their heterostructures by standard MBE technique.

**Acknowledgements:** This work was supported by National Science Foundation and Ministry of Science and Technology of China. XLQ and SCZ are supported by the Department of Energy, Office of Basic Energy Sciences, Division of Materials Sciences and Engineering, under contract DE-AC02-76SF00515. The STM topographic images were processed by WSxM software (www.nanotec.es).

# References

1 H. J. Zhang, C. X. Liu, X. L. Qi, X. Dai, Z. Fang, and S. C. Zhang, Nature Phys. **5**, 438 (2009).

2 Y. Xia, D. Qian, D. Hsieh, L. Wray, A. Pal, H. Lin, A. Bansil, D. Grauer, Y. S. Hor, R. J. Cava, and M. Z. Hasan, Nature Phys. **5**, 398 (2009).

3 D. Hsieh, Y. Xia, D. Qian, L. Wray, J. H. Dil, F. Meier, J. Osterwalder, L. Patthey, J. G. Checkelsky, N. P. Ong, A. V. Fedorov, H. Lin, A. Bansil, D. Grauer, Y. S. Cava, and M. Z. Hasan, Nature (London) **460**, 1101 (2009).

4 Y. Zhang, K. He, C. Z. Chang, C. L. Song, L. L. Wang, X. Chen, J. J. Jia, Z, Fang, X. Dai, W. Y. Shan, S. Q. Shen, Q. Niu, X. L. Qi, S. C. Zhang, X. C. Ma, and Q. K. Xue, Nature Phys. (DOI:10.1038/NPHYS1689, June 13, 2010).

5 Y. L. Chen, J. G. Analytis, J. H. Chu, Z. K. Liu, S. K. Mo, X. L. Qi, H. J. Zhang, D. H. Lu, X. Dai, Z. Fang, S. C. Zhang, I. R. Fisher, Z. Hussain, and Z. X. Shen, Science **325**, 178 (2009).

6 Y. Y. Li, G. Wang, X. G. Zhu, M. H. Liu, C. Ye, X. Chen, Y. Y. Wang, K. He, L. L.Wang, X. C. Ma, H. J. Zhang, X. Dai, Z. Fang, X. C. Xie, Y. Liu, X. Q. Qi, J. F. jia, S. C. Zhang, and Q. K. Xue, Arxiv: 0912.5054 (2009).

7 P. Roushan, J. Seo, C. V. Parker, Y. S. Hor, D. hsieh, D. Qian, A. Richardella, M. Z. Hasan, R. J. Cava, and A. Yazdani, Nature (London) **460**, 1106 (2009).

8 T. Zhang, P. Cheng, X. Chen, J. J. Jia, X. C. Ma, K. He, L. L. Wang, H. J. Zhang, X. Dai, Z. Fang, X. C. Xie, and Q. K. Xue, Phys. Rev. Lett. **103**, 266803 (2009).

9 X. L. Qi and S. C. Zhang, Phys. Today **63**, 33 (2010).

10 J. E. Moore, Nature (London) **464**, 194 (2010).

11 M. Z. Hasan and C. L. Kane, Arxiv:1002.3895 (2010).

12 Y. S. Hor, A. Richardella, P. Roushan, Y. Xia, J. G. Checkelsky, A. Yazdani, M. Z. Hasan, N. P. Ong, and R. J. Cava, Phys. Rev. B **79**, 195208 (2009).

13 S. Urazhdin, D. Bilc, S. H. Tessmer, and S. D. Mahanti, Phys. Rev. B **66**, 161306(R) (2002).

14 S. Urazhdin, D. Bilc, S. D. Mahanti, and S. H. Tessmer, Phys. Rev. B **69**, 085313 (2004).

15 A. K. Geim, and K. S. Novoselov, Nature Mater. **6**, 183 (2007).

16 J. Hass, W. A. D. Heer, and E. H. Conrad, J. Phys.: Condens. Matter. **20**, 323206 (2008).

17 X.-C. Ma, P. Jiang, Y. Qi, J.-F. Jia, Y. Yang, W.-H. Duan, W.-X. Li, X. Bao, S. B. Zhang, and Q.-K. Xue, Proc. Natl. Acad. Sci. U.S.A. **104**, 9204 (2007).

18 Q. K. Xue, T. Hashizume, and T. Sakurai, Prog. Surf. Sci. **56**, 1 (1997).

19 H. Okamoto, and L. E. Tanner, Bi-Se (Bismuth-Selenium) Phase Diagram, ASM (1986).

20 G. H. Zhang, H. J. Qin, J. Teng, J. D. Guo, Q. L. Guo, X. Dai, Z, Fang, and K. H. Wu, Appl. Phys. Lett. **95**, 053114 (2009).

21 R. D. Bringans and M. A. Olmstead, Phys. Rev. B **39**, 12985 (1989).

**Figure captions**

FIG. 1. (a) RHEED intensity of the (0, 0) diffraction versus growth time. (b) RHEED pattern along Γ-K direction and (c) STM image of the 26 QL $Bi_2Se_3$ film acquired at a sample bias ($V_s$) of 1.0V. (d) STM image ($V_s$=10 mV) showing the Se vacancies (the blue arrows).

FIG. 2. (a) ARPES intensity map of the 26 QL film along the Γ-K direction. The green dotted line indicates the Fermi level, and the red ones the topological surface states. (b) thickness-dependent surface Dirac point of $Bi_2Se_3$ films on graphene (black curve) and Si (red curve) substrates.

FIG. 3. (a) Large scale ($V_s$=4.0 V) and (b) atomic resolution STM images ($V_s$=10 meV) of double-layer graphene on SiC(0001) substrate. (c) STM image ($V_s$=1.0V) recorded after deposition of 1.5 QL $Bi_2Se_3$ on graphene. (d) High-resolution STM image ($V_s$=10 meV) of 1 QL $Bi_2Se_3$ film. The bright spot corresponds to a topmost surface Se atom. (e) *dI/dV* mapping of 1 QL $Bi_2Se_3$ film at $V_s$=-0.35V, revealing a Moiré pattern with a period of 7.10 nm. (f) Simulated Moiré pattern by assuming a fully relaxed $Bi_2Se_3$ on graphene.

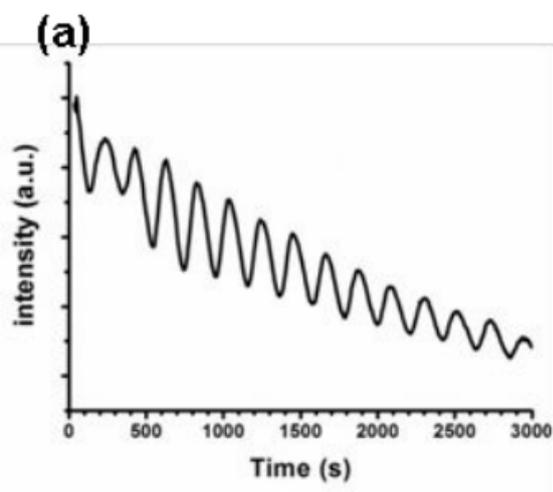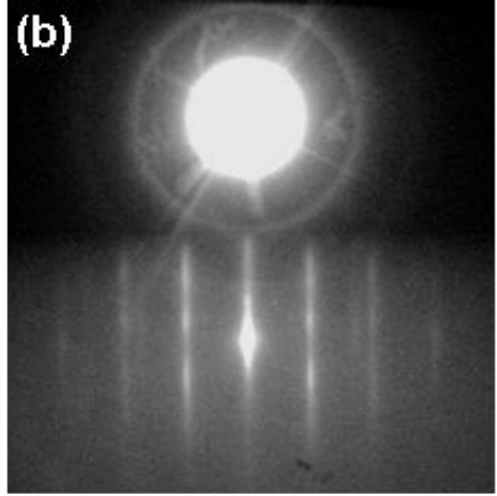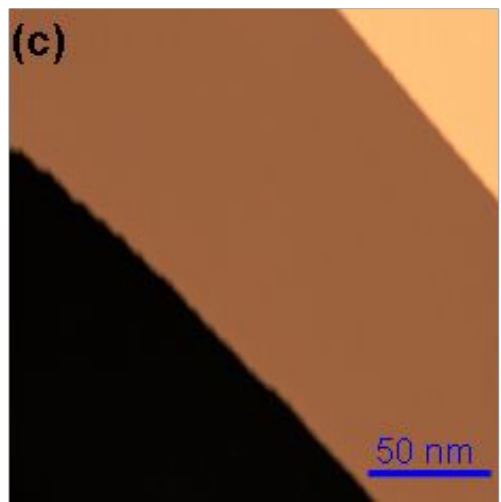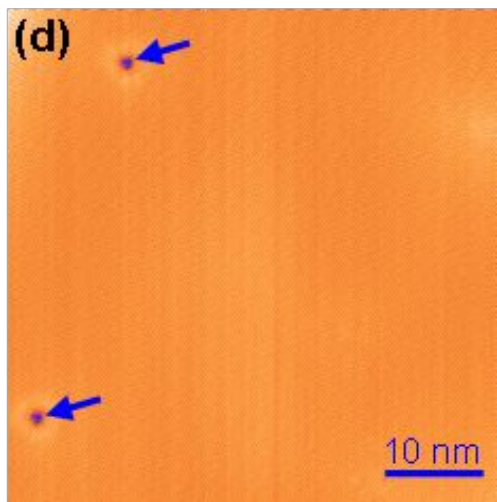

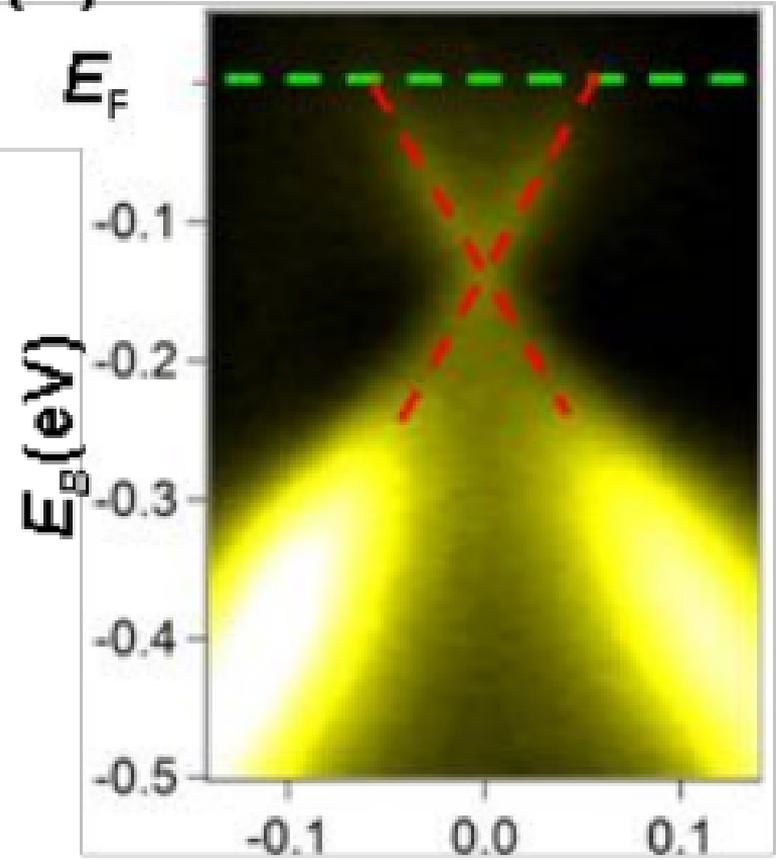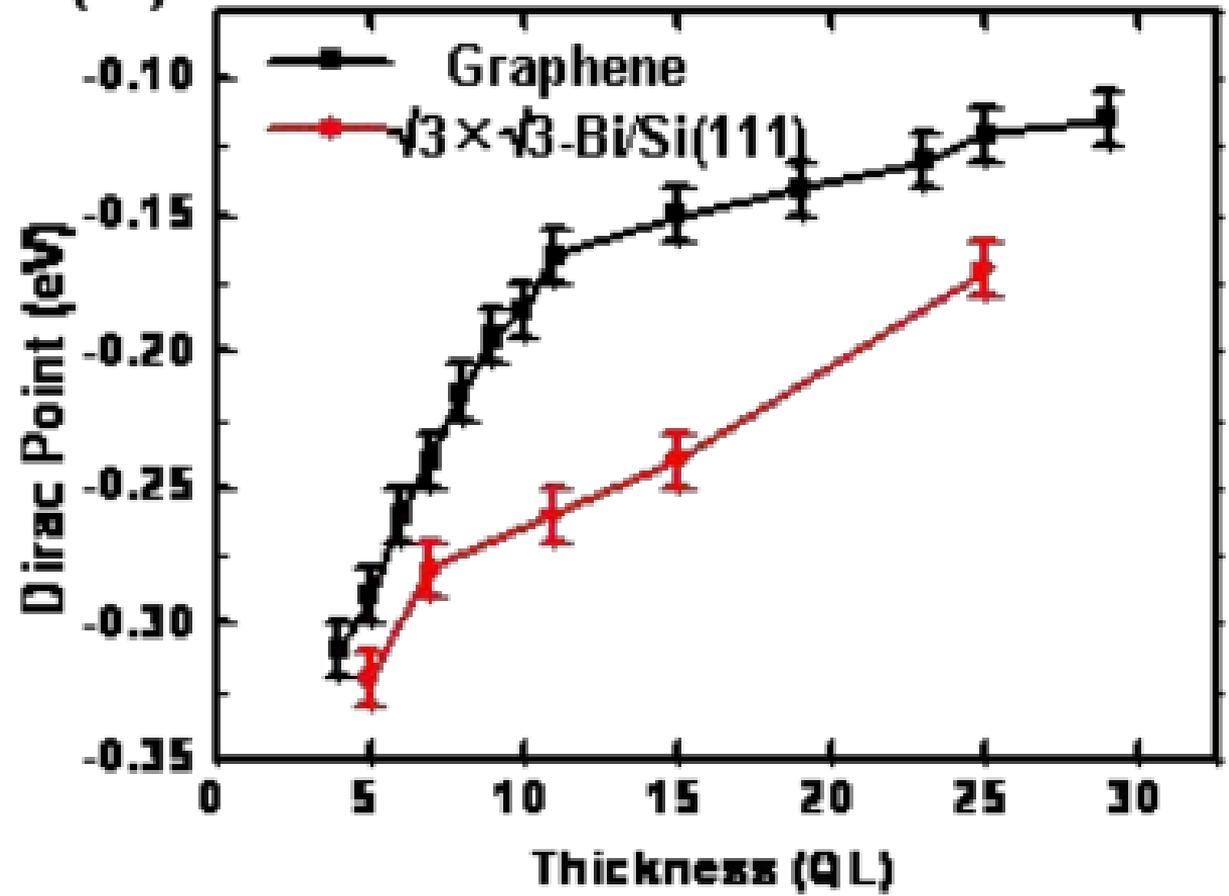

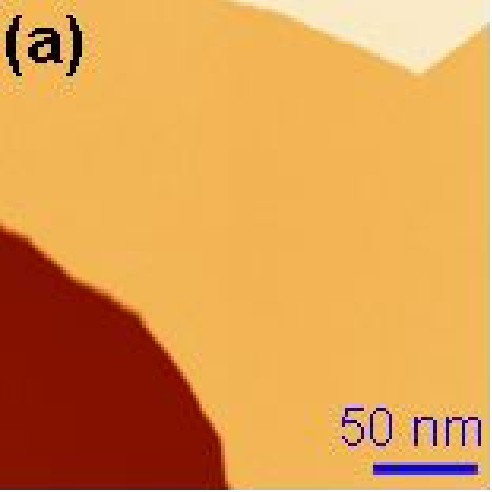 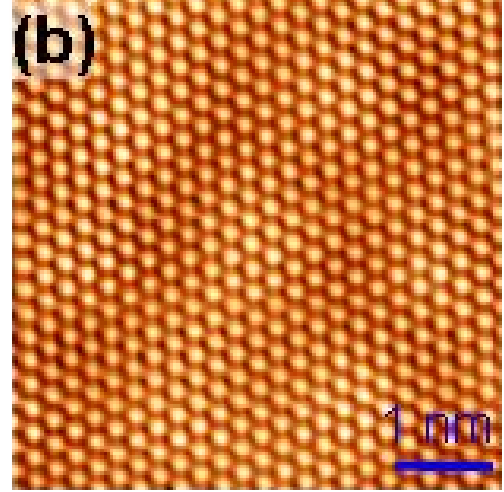 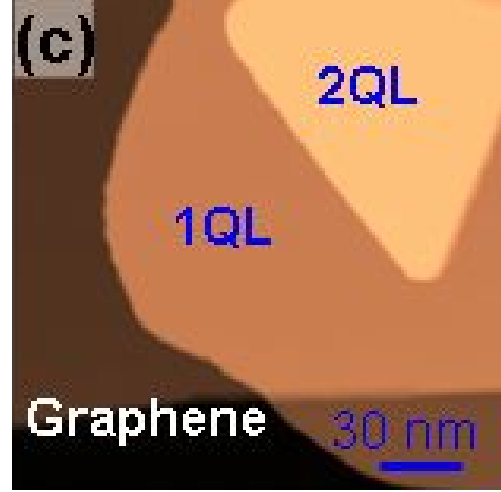 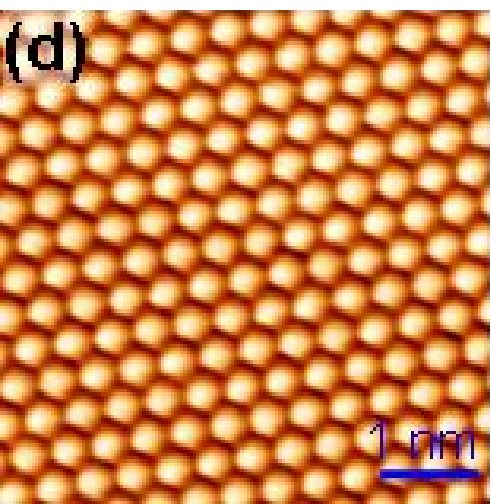 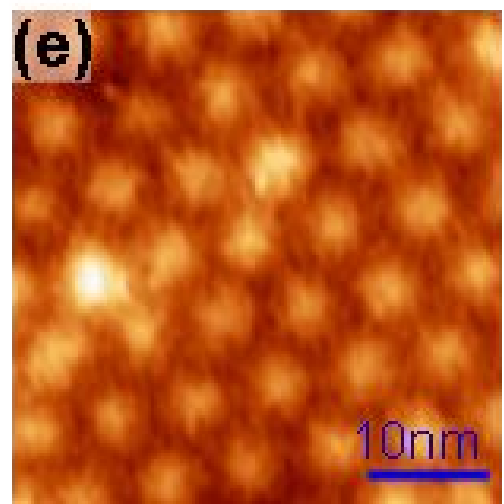 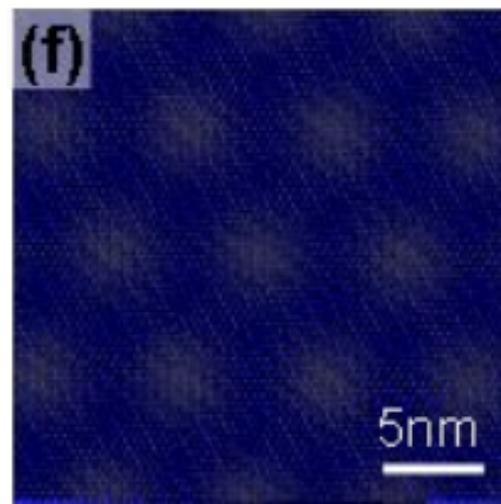